\documentclass[reprint,superscriptaddress,amsmath,amssymb,aps,prb]{revtex4-2}

\usepackage{graphicx}
\usepackage{dcolumn}
\usepackage{bm}
\usepackage[colorlinks=true, linkcolor=blue, citecolor=blue, urlcolor=blue]{hyperref}

\begin{document}

\title{Electronic structure and two-orbital model of the quadlayer La$_5$Ni$_4$O$_{13}$}

\author{Jian-Xiang Sun}
\affiliation{Guangdong Provincial Key Laboratory of Magnetoelectric Physics and Devices, State Key Laboratory of Optoelectronic Materials and Technologies, Institute of Neutron Science and Technology, School of Physics, Sun Yat-sen University, Guangzhou 510275, China}

\author{Haokan Xiao}
\affiliation{Guangdong Provincial Key Laboratory of Magnetoelectric Physics and Devices, State Key Laboratory of Optoelectronic Materials and Technologies, Institute of Neutron Science and Technology, School of Physics, Sun Yat-sen University, Guangzhou 510275, China}

\author{Cui-Qun Chen}
\email{chencq7@mail2.sysu.edu.cn} 
\affiliation{Guangdong Provincial Key Laboratory of Magnetoelectric Physics and Devices, State Key Laboratory of Optoelectronic Materials and Technologies, Institute of Neutron Science and Technology, School of Physics, Sun Yat-sen University, Guangzhou 510275, China}

\author{Dao-Xin Yao}
\email{yaodaox@mail.sysu.edu.cn}
\affiliation{Guangdong Provincial Key Laboratory of Magnetoelectric Physics and Devices, State Key Laboratory of Optoelectronic Materials and Technologies, Institute of Neutron Science and Technology, School of Physics, Sun Yat-sen University, Guangzhou 510275, China}

\date{\today}

\begin{abstract}

The discovery of pressure-induced superconductivity in Ruddlesden--Popper (RP) nickelates has stimulated extensive interest in high-T$_c$ superconductors. Here, we systematically study the electronic properties of the quadlayer RP nickelate La$_5$Ni$_4$O$_{13}$ under ambient pressure, 5\% isotropic compressive strain, and 4\% $c$-axis uniaxial strain using density functional theory (DFT) and random phase approximation (RPA) calculations. DFT calculations show that isotropic strain broadens the Ni-$e_g$ bands and induces charge transfer from O-$p$ to Ni-$d$ orbitals, whereas $c$-axis uniaxial strain selectively shifts the $d_{z^2}$-derived bonding1 band upward while leaving the $d_{x^2-y^2}$ dispersion nearly unchanged.
From Wannier downfolding, we construct a quadlayer two-orbital model that reproduces the low-energy Ni-$e_g$ bands. Our model reveals that under ambient pressure and 5\% isotropic strain, the Fermi surface consists of two electron pockets ($\alpha$ and $\delta$) and three hole pockets ($\beta$, $\beta^{\prime}$, and $\beta^{\prime \prime}$), while under uniaxial strain, a $\gamma$ hole pocket with $d_{z^2}$ orbital character emerges. RPA calculations reveal that the leading spin response shifts from $\mathbf{q}\approx(2\pi/3,2\pi/3)$ at ambient pressure to $\mathbf{q}\approx(\pi,\pi)$  under both strain conditions and is enhanced under $c$-axis compression. These results suggest that $c$-axis compression may provide a favorable route to superconductivity in the quadlayer nickelate analogous to that in bilayer and trilayer nickelates.
\end{abstract}

\maketitle

\section{Introduction} 
 
The discovery of superconductivity in the infinite-layer nickelates has drawn considerable attention to unconventional superconductivity in nickel-based materials \cite{Li2019NdNiO2, Nomura2022Review}. More recently, the Ruddlesden--Popper (RP) nickelates $\mathrm{La}_{n+1}\mathrm{Ni}_n\mathrm{O}_{3n+1}$ have emerged as an important new family \cite{Sun2023La3Ni2O7, Zhu2024NatureLa4Ni3O10, Li2024La4Ni3O10, Zhang2025Trends, Puphal2026Review}. In these RP systems, pressure suppresses structural distortions and stabilizes higher-symmetry structures \cite{Wang2024JACSStructure, Geisler2024npj, Li2024SCPMA, Chen2024Trilayer, LaBollita2024Trilayer}. Notably, the emergence of superconductivity is closely tied to these pressure-induced changes in crystal and electronic structures \cite{Wang2024Review}. The superconducting transition temperature (T$_c$) approaches $80$~K in the bilayer nickelate $\mathrm{La}_3\mathrm{Ni}_2\mathrm{O}_7$ ($n=2$) above $\sim 14$~GPa \cite{Sun2023La3Ni2O7, Wang2024PRX, Hou2023CPL}, and reaches $20$~K in the trilayer $\mathrm{La}_4\mathrm{Ni}_3\mathrm{O}_{10}$ ($n=3$) at approximately $43$~GPa \cite{Zhu2024NatureLa4Ni3O10, Li2024La4Ni3O10}. These observations naturally raise the question of how the electronic structure and superconducting properties evolve in higher-order members of the RP series \cite{Li2020APLMater, Jung2022HigherOrder, Zhang2025Trends}. 
 
Pressure-induced modifications of the electronic structure in RP nickelates involve bandwidth broadening, enhanced interlayer coupling, and charge redistribution \cite{Christiansson2023PRL, Zhang2023BilayerMagnetic, Chen2024Trilayer, Li2024SCPMA, Leonov2024PRB, Wu2024SCPMA, LaBollita2024Trilayer}. In particular, pressure-induced upward shifts of the Ni-$d_{z^2}$-derived bands toward the Fermi level have been widely discussed in the bilayer and trilayer RP nickelates \cite{Christiansson2023PRL, Zhang2023BilayerMagnetic, Li2024SCPMA, Sakakibara2024PRB, Chen2024Trilayer, LaBollita2024Trilayer}. In these systems, the resulting $d_{z^2}$-derived hole pocket near the Brillouin-zone corner has been suggested to play an important role in the superconducting pairing mechanism \cite{Luo2023Bilayer, Shilenko2023PRB, Chen2024Trilayer, LaBollita2024Trilayer, Yang2024TrilayerModel, Zhang2024TrilayerRPA, Luo2024BilayerTJ}. Therefore, the evolution of such $d_{z^2}$ states provides a useful perspective for comparing different members of the RP series \cite{Pardo2011PRB, Li2017ARPES, Jung2022HigherOrder, Zhao2025Strain, Zhang2025Trends}. 


Experimentally, higher-order RP nickelates with $n=4$ and $5$ have also been stabilized in thin-film form \cite{Li2020APLMater, Shi2026RP}. Extending these investigations to higher-order RP members ($n \ge 4$) is therefore important for understanding the evolution of the RP nickelate series \cite{Li2020APLMater, Jung2022HigherOrder, Zhao2025Strain, Zhang2025Trends}. 
Previous theoretical studies of $\mathrm{La}_5\mathrm{Ni}_4\mathrm{O}_{13}$ have reported multiple Fermi surface sheets and a molecular-subband structure arising from the four coupled Ni-$d_{z^2}$ orbitals \cite{Jung2022HigherOrder, Zhao2025Strain, Zhang2025Trends}. Random phase approximation (RPA) calculations have also found nearly degenerate $d_{x^2-y^2}$-wave and $s^{\pm}$-wave leading pairing states, while density functional theory (DFT) calculations revealed the absence of a small $d_{z^2}$-derived hole pocket under high-pressure conditions \cite{Zhang2025Trends}. However, a systematic comparison of the effects of isotropic and uniaxial strain on the quadlayer electronic properties remains lacking \cite{Zhao2025Strain, Zhang2025Trends}. 
 
In this work, we study the electronic structure of quadlayer $\mathrm{La}_5\mathrm{Ni}_4\mathrm{O}_{13}$ under ambient pressure, $5\%$ compressive isotropic strain, and $4\%$ $c$-axis uniaxial strain where the $\gamma$ pocket emerges. DFT calculations show that $5\%$ isotropic strain broadens the Ni-$e_g$ bands and induces a redistribution of the Ni-$d$ and O-$p$ orbital occupancies, whereas $4\%$ $c$-axis uniaxial strain selectively shifts the $d_{z^2}$-derived subbands upward. Furthermore, we establish an effective quadlayer two-orbital tight-binding model through Wannier downfolding, which reproduces the main low-energy Ni-$e_g$ dispersions and captures the evolution of the Fermi surfaces across different strains. In particular, the $d_{z^2}$-derived $\gamma$ hole pocket remains absent under $5\%$ isotropic strain but emerges under $4\%$ uniaxial strain. Finally, we calculate the static RPA spin susceptibility. The largest eigenvalue peaks near $(2\pi/3,2\pi/3)$ at ambient pressure and at $M$ under both strain conditions. The $M$-point peak under isotropic strain suggests a possible contribution from virtual scattering involving the incipient $\gamma$ band, similar to that proposed for trilayer $\mathrm{La}_4\mathrm{Ni}_3\mathrm{O}_{10}$ \cite{Zhang2024PressurizedTrilayer}. Under $c$-axis compression, the $\gamma$ pocket appears and the spin susceptibility is enhanced. These similarities with bilayer and trilayer nickelates suggest that $c$-axis compression may provide favorable conditions for superconductivity in the quadlayer system \cite{Yang2023PRB, Liu2023PRL, Zhang2024BilayerRPA, Yang2024TrilayerModel, Zhang2024TrilayerRPA, Zhang2024PressurizedTrilayer}.

\begin{figure}[htbp]
\centering
\includegraphics[width=\columnwidth]{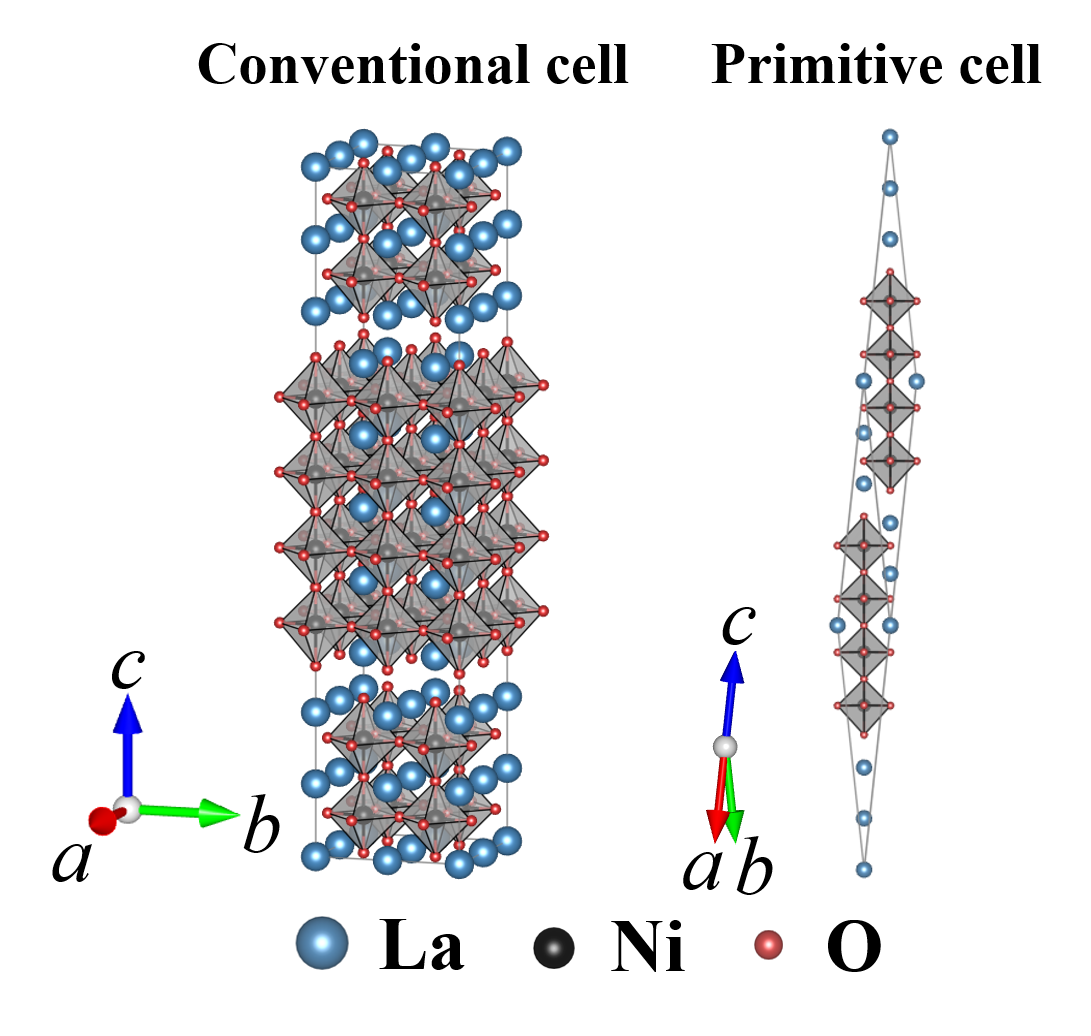}
\caption{\label{fig:structure}
Crystal structure of the high-symmetry $I4/mmm$ phase for the quadlayer $\mathrm{La}_5\mathrm{Ni}_4\mathrm{O}_{13}$. The left and right panels show the conventional and primitive cells, respectively. The La, Ni, and O atoms are denoted by blue, black, and red balls, respectively.
}
\end{figure}

\begin{figure*}[htbp]
\centering
\includegraphics[width=\textwidth]{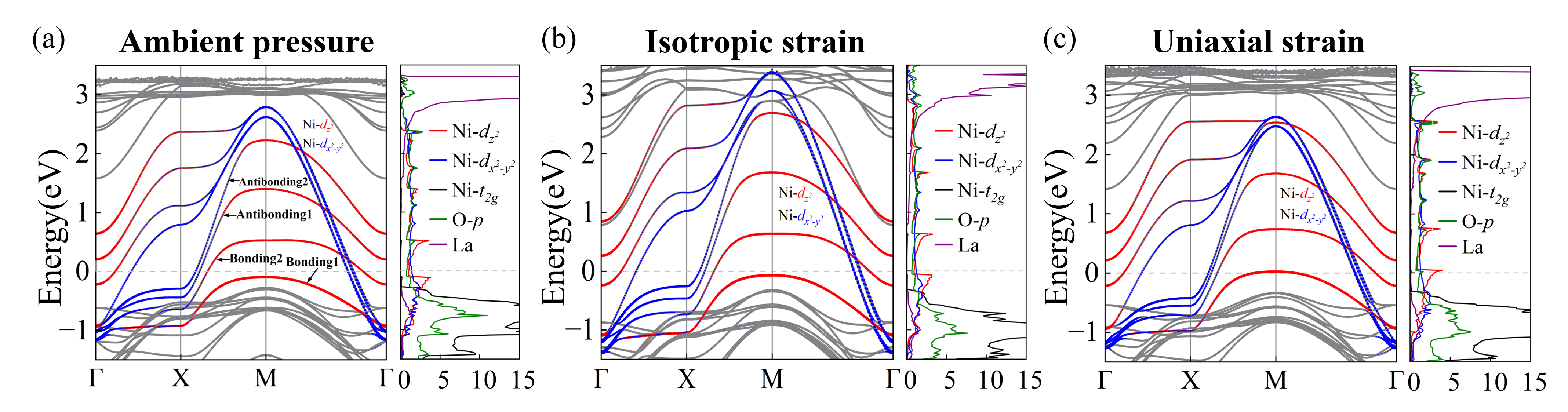}
\caption{\label{fig:dft_dos} 
(a--c) DFT-calculated electronic band structures alongside the projected density of states (DOS) under 
(a) ambient pressure, 
(b) $5\%$ isotropic strain, and 
(c) $4\%$ $c$-axis uniaxial strain.
In the band structure panels, red and blue curves represent the orbital weights of Ni-$d_{z^2}$ and Ni-$d_{x^2-y^2}$, respectively. In the DOS panels, red, blue, black, green, and purple curves correspond to Ni-$d_{z^2}$, Ni-$d_{x^2-y^2}$, Ni-$t_{2g}$, O-$p$, and $\mathrm{La}$ states, respectively.}
\end{figure*}

\section{Crystal Structure and Computational Methods}

Since bulk $\mathrm{La}_5\mathrm{Ni}_4\mathrm{O}_{13}$ has not been experimentally synthesized yet, we construct a structure with the tetragonal $I4/mmm$ space group for the following calculations based on the structures of thin-film La$_5$Ni$_4$O$_{13}$ and bulk La$_4$Ni$_3$O$_{10}$ \cite{Li2020APLMater,Jung2022HigherOrder,Li2017ARPES,Puggioni2018La4Ni3O10, Shi2026RP}. 
As illustrated in the left panel of Fig.~\ref{fig:structure}, each Ni ion is coordinated by six oxygen atoms, forming a $\mathrm{NiO}_6$ octahedron \cite{Greenblatt1997, Jung2022HigherOrder}. The corner-sharing octahedra form four $\mathrm{NiO}_2$ layers, which are coupled through apical oxygen atoms via interlayer Ni--O--Ni bonds along the $c$ axis. Neighboring quadlayer perovskite blocks are separated by rocksalt-type La--O spacer layers \cite{Greenblatt1997, Li2020APLMater, Jung2022HigherOrder}. A primitive unit cell containing four Ni atoms (one formula unit) was employed in all calculations (right panel of Fig.~\ref{fig:structure}).

DFT calculations were performed using the Vienna \textit{ab initio} Simulation Package (VASP) \cite{Kresse1996VASP, Kresse1996VASP_PRB}. The exchange-correlation functional was treated within the generalized gradient approximation (GGA) using the Perdew--Burke--Ernzerhof (PBE) functional \cite{Perdew1996PBE}, together with the projector augmented-wave (PAW) method \cite{Blochl1994PAW, Kresse1999PAW}. A plane-wave cutoff energy of $500$~eV was used throughout the calculations. The electronic energy convergence criterion was set to $10^{-6}$~eV, and the internal atomic coordinates were fully relaxed until the residual forces on each atom were smaller than $0.001$~eV/\AA. The lattice parameters were fixed during the structural relaxations. The Brillouin zone was sampled using a $15\times15\times15$ Monkhorst--Pack $k$-point mesh \cite{Monkhorst1976}.


To construct low-energy tight-binding models, maximally localized Wannier functions were generated for the Ni-$e_g$ orbitals using the WANNIER90 package \cite{Marzari1997Wannier, Souza2001Wannier, Mostofi2008Wannier90, Pizzi2020Wannier90}. The resulting tight-binding models reproduce the main features of the low-energy Ni-$e_g$ band dispersions around the Fermi level obtained from DFT calculations.

\section{Results and Discussion}

\subsection{Electronic structure and strain-driven evolution}

Figure~\ref{fig:dft_dos} presents the DFT-calculated electronic band structures and the projected density of states (DOS) for $\mathrm{La}_5\mathrm{Ni}_4\mathrm{O}_{13}$ under ambient pressure, $5\%$ compressive isotropic strain, and $4\%$ $c$-axis compressive uniaxial strain, respectively. 
Under the NiO$_6$ octahedral crystal field, the Ni-$d$ orbitals split into $e_g$ and t$_{2g}$ sectors. Similar to the bilayer La$_3$Ni$_2$O$_7$ and trilayer La$_4$Ni$_3$O$_{10}$, the low-energy bands near the Fermi level are predominantly derived from Ni-$e_g$ orbitals, while the nearly fully occupied Ni-$t_{2g}$ states lie well below the Fermi level \cite{Luo2023Bilayer, Chen2024Trilayer}. The $d_{z^2}$ orbitals couple through the apical oxygen atoms between $\mathrm{NiO}_2$ planes, splitting into two bonding bands (bonding1 and bonding2) and two antibonding bands (antibonding1 and antibonding2). Among these, only the bonding1 band resides below the Fermi level, while the others lie above it.

In bulk bilayer and trilayer RP nickelates, pressure is a key factor for superconductivity \cite{Sun2023La3Ni2O7, Zhu2024NatureLa4Ni3O10}. To simulate the effect of pressure, we performed calculations under 5\% isotropic strain, which is equivalent to a pressure of 36 GPa as DFT reveals. Under 5\% isotropic strain, the overall Ni-$e_g$ bandwidth increases from 3.95 eV at ambient pressure to 4.77 eV. Notably, the bonding1 band does not cross the Fermi level under pressure and therefore fails to form a $\gamma$ pocket, in contrast to the pressure-induced Lifshitz transitions observed in the bilayer La$_3$Ni$_2$O$_7$ and trilayer La$_4$Ni$_3$O$_{10}$ \cite{Sun2023La3Ni2O7,Christiansson2023PRL,Zhang2023BilayerMagnetic,Chen2024Trilayer,Li2024SCPMA}. Calculations at 8\% and 10\% strain (corresponding to 73 GPa and 107 GPa, respectively) show similar behavior, as presented in the Supplementary Materials. These results suggest that achieving superconductivity in La$_5$Ni$_4$O$_{13}$ through hydrostatic pressure alone appears to be difficult, at least within the framework where the $\gamma$ pocket is considered essential for superconductivity.

Uniaxial strain has previously been proposed as an effective route to induce the $\gamma$ pocket \cite{Huo2025Strain}. We therefore calculated the electronic structure under uniaxial strain. At 3\% compressive strain along the $c$ axis, the bonding1 band remains below the Fermi level (see Supplementary Materials). However, when the $c$-axis compressive strain exceeds 4\% (corresponding to $\sim$7 GPa), the bonding1 band crosses the Fermi level and forms a $\gamma$ hole pocket near the M point. Meanwhile, the DOS of the Ni-$d_{z^2}$ orbital exhibits a peak at the Fermi level, which is favorable for superconductivity. Furthermore, the effect of uniaxial strain is predominantly manifested in the $d_{z^2}$ orbital, while the $d_{x^2-y^2}$ orbital remains nearly unaffected with the bandwidth nearly the same as at ambient pressure. Uniaxial strain reduces the interlayer distance and thereby enhances the interlayer $d_{z^2}$ hopping, increasing the energy gap between the bonding and antibonding bands. Consequently, the $d_{z^2}$ bandwidth increases from 3.29 eV at ambient pressure to 3.53 eV, becoming almost comparable to that of the $d_{x^2-y^2}$ orbital.

\begin{table}[htbp]
\caption{\label{tab:occupation} DFT-calculated Ni-$d$ and O-$p$ orbital occupancies for the Inner and Outer layers of $\mathrm{La}_5\mathrm{Ni}_4\mathrm{O}_{13}$ under ambient pressure, $5\%$ compressive isotropic strain, and $4\%$ $c$-axis uniaxial strain. Here, Ni-$t_{2g}$ denotes the total occupation of the three $t_{2g}$ orbitals. O-$p_{x/y}$ denotes the in-plane O-$p_x$ or O-$p_y$ orbital oriented along the Ni--O bond. For O-$p_z$, the Inner and Outer rows correspond to apical oxygen sites within the quadlayer block and on its two outer boundaries, respectively.}
\renewcommand{\arraystretch}{1.2}
\resizebox{\columnwidth}{!}{%
\begin{tabular}{c c c c c c c}
\hline\hline
Condition & Layer & Ni-$t_{2g}$ & Ni-$d_{z^2}$ & Ni-$d_{x^2-y^2}$ & O-$p_z$ & O-$p_{x/y}$ \\
\hline
Ambient pressure    & Inner & 5.799 & 1.330 & 1.221 & 1.574 & 1.644 \\
      & Outer & 5.745 & 1.306 & 1.275 & 1.820 & 1.585 \\
Isotropic strain & Inner & 5.876 & 1.310 & 1.256 & 1.547 & 1.638 \\
      & Outer & 5.813 & 1.337 & 1.325 & 1.815 & 1.580 \\
Uniaxial strain    & Inner & 5.795 & 1.323 & 1.277 & 1.585 & 1.545 \\
      & Outer & 5.757 & 1.288 & 1.320 & 1.809 & 1.580 \\
\hline\hline
\end{tabular}%
}
\end{table}

\begin{figure}[htbp]
\centering
\includegraphics[width=\columnwidth]{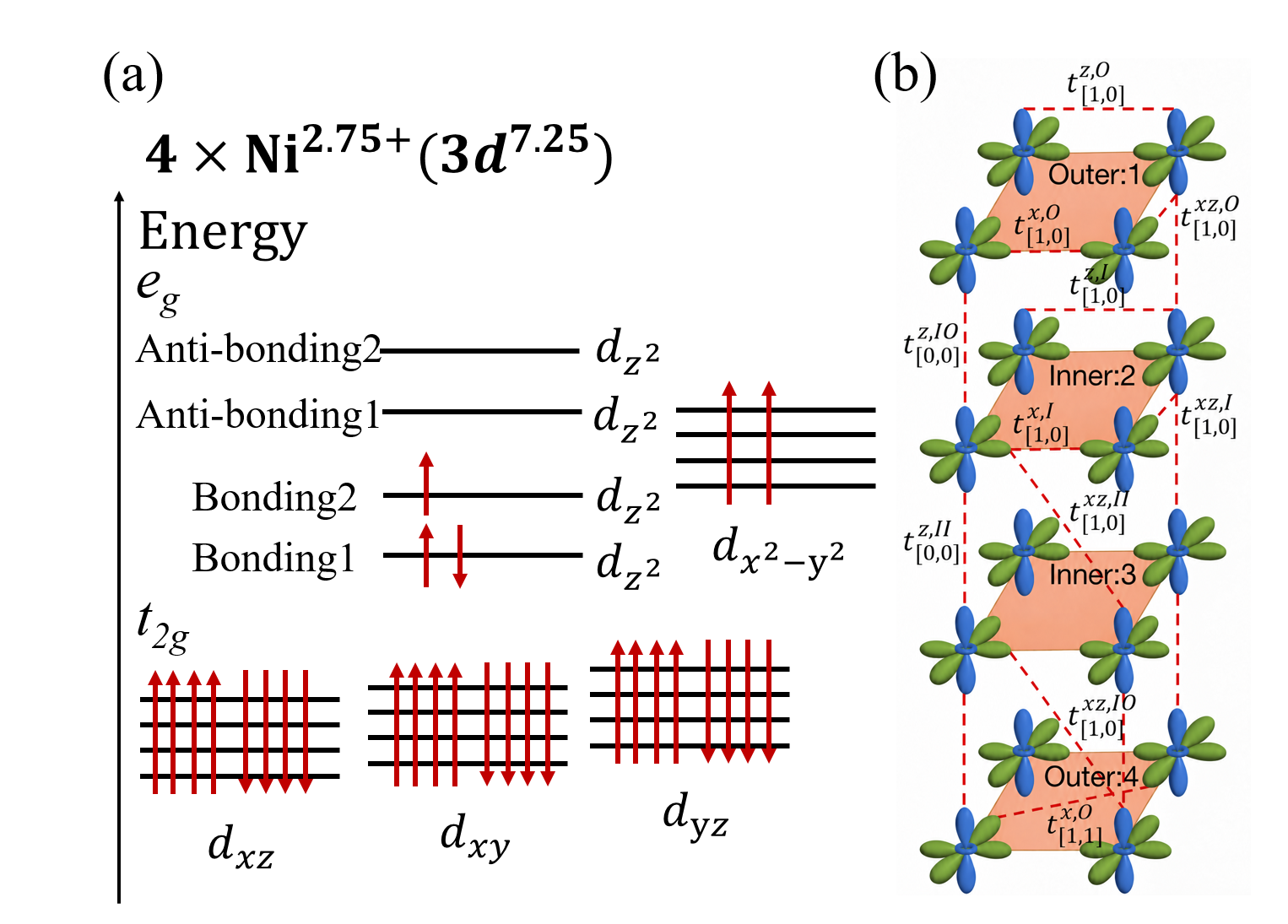}
\caption{\label{fig:atomic_model}
(a) Schematic energy-level diagram and nominal electronic configuration for $\mathrm{La}_5\mathrm{Ni}_4\mathrm{O}_{13}$. (b) Schematic of the quadlayer lattice. The blue and green shapes indicate Ni-$d_{z^2}$ and $d_{x^2-y^2}$ orbitals, respectively. Some of the hoppings of the quadlayer two-orbital model are drawn, whose values can be found in Table~\ref{tab:hopping}.}
\end{figure}

To investigate strain-induced charge redistribution, we move on to the Ni-$d$ and O-$p$ orbital occupancies. The energy-level diagram and nominal configuration are illustrated schematically in Fig.~\ref{fig:atomic_model}(a). The nominal Ni valence of +2.75 corresponds to a $3d^{7.25}$ configuration, which, with a fully occupied $t_{2g}$ manifold, leaves five $e_g$ electrons per four-Ni primitive cell. 
The DFT-calculated Ni-$d$ and O-$p$ orbital occupancies are summarized in Table~\ref{tab:occupation}. Note that the DFT-calculated occupancies of Ni-$e_g$ orbitals are larger than the nominal ones, reflecting orbital hybridization. 
At ambient pressure, the Ni-$t_{2g}$ orbitals are nearly filled, with an average occupation of 5.77 electrons per Ni, while the $e_g$ sector contains 2.57 electrons per Ni, with $d_{z^2}$ occupancies slightly larger than those of the $d_{x^2-y^2}$ orbital. 

Under 5\% isotropic strain, the total Ni-$3d$ occupation increases from 8.34 to 8.46 per Ni, with both $e_g$  and $t_{2g}$ sectors gaining charge, while O-$p$ occupations decrease slightly in both layers. This indicates a net O-$p$ to Ni-$d$ charge transfer. In contrast, 4\% $c$-axis uniaxial strain leaves the $t_{2g}$ occupation nearly unchanged and shows an orbital-dependent effect on the $e_g$ sector. The $d_{x^2-y^2}$ occupation increases by 0.05 on average, whereas $d_{z^2}$ decreases slightly by 0.01. The O-$p$ response is layer dependent, with O-$p_{x/y}$ decreasing mainly in the inner layer and O-$p_z$ exhibiting opposite trends in the two layers. Overall, uniaxial strain drives charge toward Ni-$d$ states, preferentially enhancing $d_{x^2-y^2}$ and reducing $d_{z^2}$, consistent with the upward shift of the bonding1 band.

\subsection{Quadlayer two-orbital model}

To describe the low-energy electronic structure and the strain-induced evolution, we construct an effective quadlayer two-orbital model through Wannier downfolding on the Ni-$e_g$ orbitals \cite{Luo2023Bilayer, Gu2023, Sakakibara2024PRL, Chen2024Trilayer, Yang2024TrilayerModel, Zhang2025Trends, Hu2025FilmModel, Li2026FilmInterface}. The effective two-orbital model can be written as
\begin{equation}
\mathcal{H} = \mathcal{H}_t + \mathcal{H}_U, 
\quad 
\mathcal{H}_t = \sum_{\mathbf{k}\sigma} 
\Psi_{\mathbf{k}\sigma}^\dagger 
H(\mathbf{k}) 
\Psi_{\mathbf{k}\sigma},
\end{equation}
where $\mathcal{H}_t$ denotes the noninteracting tight-binding Hamiltonian and $\mathcal{H}_U$ describes the Coulomb interactions. The local interaction term $\mathcal{H}_U$ is written in the Kanamori form \cite{Kanamori1963, Georges2013Kanamori, Chen2024Trilayer}:
\begin{equation}
\begin{aligned}
\mathcal{H}_U &= U \sum^{\alpha=1,2,3,4}_i
\left( 
n_{i\uparrow}^{x_\alpha} n_{i\downarrow}^{x_\alpha} 
+ n_{i\uparrow}^{z_\alpha} n_{i\downarrow}^{z_\alpha} 
\right) \\
&\quad + U' \sum^{\alpha=1,2,3,4}_{i\sigma} 
n_{i\sigma}^{x_\alpha} n_{i\bar{\sigma}}^{z_\alpha} 
+ (U' - J_H) \sum^{\alpha=1,2,3,4}_{i\sigma}
n_{i\sigma}^{x_\alpha} n_{i\sigma}^{z_\alpha} \\
&\quad + J_H \sum^{\alpha=1,2,3,4}_i
\Big( 
c_{ix_\alpha \uparrow}^\dagger 
c_{ix_\alpha \downarrow}^\dagger 
c_{iz_\alpha \downarrow} 
c_{iz_\alpha \uparrow} \\
&\qquad\qquad\quad 
- c_{ix_\alpha \uparrow}^\dagger 
c_{ix_\alpha \downarrow} 
c_{iz_\alpha \downarrow}^\dagger 
c_{iz_\alpha \uparrow} 
+ \mathrm{H.c.} 
\Big).
\end{aligned}
\end{equation}

The basis is defined as
\[
\Psi_{\mathbf{k}\sigma}^\dagger =
\left(
c_{\mathbf{k}x1\sigma}^\dagger,
c_{\mathbf{k}x2\sigma}^\dagger,
c_{\mathbf{k}x3\sigma}^\dagger,
c_{\mathbf{k}x4\sigma}^\dagger,
c_{\mathbf{k}z1\sigma}^\dagger,
c_{\mathbf{k}z2\sigma}^\dagger,
c_{\mathbf{k}z3\sigma}^\dagger,
c_{\mathbf{k}z4\sigma}^\dagger
\right),
\]
where $c_{\mathbf{k}x\alpha\sigma}^\dagger$ 
($c_{\mathbf{k}z\alpha\sigma}^\dagger$) denotes the creation of a $d_{x^2-y^2}$ ($d_{z^2}$) electron in the layer $\alpha$ 
($\alpha=1,2,3,4$) with momentum $\mathbf{k}$ and spin $\sigma$. 
The local density operator is defined as 
$n_{i\sigma}^{\gamma_\alpha}
= c_{i\gamma_\alpha\sigma}^\dagger c_{i\gamma_\alpha\sigma}$, 
where $\gamma=x,z$. 
In the present Kanamori parametrization, we take 
$U' = U - 2J_H$, where $U$, $U'$, and $J_H$ denote the intra-orbital Coulomb repulsion, inter-orbital Coulomb repulsion, and Hund's coupling, respectively.

The tight-binding Hamiltonian is written in block form as
\begin{equation}
H(\mathbf{k}) =
\begin{bmatrix}
H^x(\mathbf{k}) & H^{xz}(\mathbf{k}) \\
H^{xz\dagger}(\mathbf{k}) & H^z(\mathbf{k})
\end{bmatrix},
\end{equation}
where $H^x(\mathbf{k})$ and $H^z(\mathbf{k})$ are the $4\times4$ intra-orbital tight-binding matrices for the $d_{x^2-y^2}$ and $d_{z^2}$ orbitals, respectively, and $H^{xz}(\mathbf{k})$ describes the inter-orbital hybridization between them.

Owing to the mirror symmetry of the quadlayer in the $I4/mmm$ structure, 
layers 1 and 4 are symmetry equivalent, as are layers 2 and 3 
\cite{Jung2022HigherOrder}. We therefore denote layers 1 and 4 as the 
Outer ($O$) layers and layers 2 and 3 as the Inner ($I$) layers, as 
illustrated in Fig.~\ref{fig:atomic_model}(b). Accordingly, we use $II$ 
and $OO$ to denote the Inner--Inner and Outer--Outer layer couplings, 
respectively, while $IO$ and $IO'$ denote the adjacent 
(e.g., layers 1--2 or 3--4) and non-adjacent 
(e.g., layers 1--3 or 2--4) Inner--Outer layer couplings, respectively.


For the intralayer terms, the matrix elements are
defined as
$H_O^\gamma \equiv H_{11}^\gamma=H_{44}^\gamma$ and
$H_I^\gamma \equiv H_{22}^\gamma=H_{33}^\gamma$, where $\gamma$ denotes the orbital-sector index, with 
$\gamma=x,z,xz$ corresponding to the $d_{x^2-y^2}$, $d_{z^2}$, and
 the inter-orbital hybridization sectors, respectively.
The intra-orbital diagonal elements are given by
\begin{equation}
\begin{aligned}
H_{O/I}^{x} &=
\epsilon^{x,O/I}
+2t_{[1,0]}^{x,O/I}(\cos k_x+\cos k_y) \\
&\quad
+4t_{[1,1]}^{x,O/I}\cos k_x\cos k_y \\
&\quad
+2t_{[2,0]}^{x,O/I}(\cos 2k_x+\cos 2k_y), \\
H_{O/I}^{z} &=
\epsilon^{z,O/I}
+2t_{[1,0]}^{z,O/I}(\cos k_x+\cos k_y) \\
&\quad
+4t_{[1,1]}^{z,O/I}\cos k_x\cos k_y .
\end{aligned}
\end{equation}

The intralayer inter-orbital hybridization between the $d_{x^2-y^2}$ and $d_{z^2}$ orbitals is expressed as
\begin{equation}
H_{O/I}^{xz}
=
2t_{[1,0]}^{xz,O/I}(\cos k_x-\cos k_y)
+2t_{[2,0]}^{xz,O/I}(\cos 2k_x-\cos 2k_y).
\end{equation}

For the interlayer sector, the matrix elements are defined as
$H_{IO}^{\gamma}\equiv H_{12}^{\gamma}=H_{34}^{\gamma}$ and
$H_{II}^{\gamma}\equiv H_{23}^{\gamma}$ for $\gamma=z,xz$,
while
$H_{OO}^{z}\equiv H_{14}^{z}$ and
$H_{IO'}^{z}\equiv H_{13}^{z}=H_{24}^{z}$,
with the Hermitian-conjugate matrix elements implied. The interlayer $d_{z^2}$--$d_{z^2}$ hopping terms are given by
\begin{equation}
\begin{aligned}
H_{IO}^z
&=t_{[0,0]}^{z,IO}
+2t_{[1,0]}^{z,IO}(\cos k_x+\cos k_y), \\
H_{II}^z
&=t_{[0,0]}^{z,II}
+2t_{[1,0]}^{z,II}(\cos k_x+\cos k_y), \\
H_{OO}^z
&=t_{[0,0]}^{z,OO}, \\
H_{IO'}^z
&=t_{[0,0]}^{z,IO'} .
\end{aligned}
\end{equation}

The corresponding interlayer inter-orbital hybridizations are
\begin{equation}
\begin{aligned}
H_{IO}^{xz}
&=2t_{[1,0]}^{xz,IO}(\cos k_x-\cos k_y), \\
H_{II}^{xz}
&=2t_{[1,0]}^{xz,II}(\cos k_x-\cos k_y).
\end{aligned}
\end{equation}

\begin{table*}[htbp]
\caption{\label{tab:hopping} Tight-binding parameters (in eV) for the quadlayer two-orbital model under ambient pressure (AP), $5\%$ compressive isotropic strain (IS), and $4\%$ $c$-axis uniaxial strain (US). $I$ and $O$ denote the Inner and Outer layers, respectively.}
\renewcommand{\arraystretch}{1.2}
\resizebox{\textwidth}{!}{%
\begin{tabular}{c rrr c rrr c rrr}
\hline\hline
\multicolumn{4}{c}{Inner Layer ($I$)} & \multicolumn{4}{c}{Outer Layer ($O$)} & \multicolumn{4}{c}{Interlayer} \\
Parameter & AP & IS & US & Parameter & AP & IS & US & Parameter & AP & IS & US \\
\hline
$\epsilon^{x,I}$    & $0.895$  & $1.107$  & $0.773$  &
$\epsilon^{x,O}$    & $0.703$  & $0.829$  & $0.608$  &
$t_{[0,0]}^{z,II}$  & $-0.544$ & $-0.722$ & $-0.639$ \\

$\epsilon^{z,I}$    & $0.792$  & $0.985$  & $0.960$  &
$\epsilon^{z,O}$    & $0.466$  & $0.502$  & $0.650$  &
$t_{[1,0]}^{z,II}$  & $0.017$  & $0.016$  & $0.029$ \\

$t_{[1,0]}^{x,I}$   & $-0.462$ & $-0.548$ & $-0.459$ &
$t_{[1,0]}^{x,O}$   & $-0.461$ & $-0.541$ & $-0.460$ &
$t_{[0,0]}^{z,IO}$  & $-0.543$ & $-0.646$ & $-0.612$ \\

$t_{[1,1]}^{x,I}$   & $0.063$  & $0.089$  & $0.077$  &
$t_{[1,1]}^{x,O}$   & $0.073$  & $0.086$  & $0.083$  &
$t_{[1,0]}^{z,IO}$  & $0.040$  & $0.031$  & $0.040$ \\

$t_{[2,0]}^{x,I}$   & $-0.064$ & $-0.084$ & $-0.077$ &
$t_{[2,0]}^{x,O}$   & $-0.065$ & $-0.076$ & $-0.073$ &
$t_{[0,0]}^{z,OO}$  & $-0.087$ & $-0.082$ & $-0.073$ \\

$t_{[1,0]}^{z,I}$   & $-0.135$ & $-0.195$ & $-0.193$ &
$t_{[1,0]}^{z,O}$   & $-0.098$ & $-0.105$ & $-0.126$ &
$t_{[0,0]}^{z,IO'}$ & $-0.073$ & $-0.105$ & $-0.076$ \\

$t_{[1,1]}^{z,I}$   & $-0.005$ & $-0.030$ & $-0.041$ &
$t_{[1,1]}^{z,O}$   & $-0.032$ & $-0.021$ & $-0.035$ &
$t_{[1,0]}^{xz,IO}$ & $0.042$  & $0.043$  & $0.030$ \\

$t_{[1,0]}^{xz,I}$  & $-0.238$ & $-0.267$ & $-0.256$ &
$t_{[1,0]}^{xz,O}$  & $-0.244$ & $-0.287$ & $-0.259$ &
$t_{[1,0]}^{xz,II}$ & $0.037$  & $0.040$  & $0.029$ \\

$t_{[2,0]}^{xz,I}$  & $-0.037$ & $-0.039$ & $-0.034$ &
$t_{[2,0]}^{xz,O}$  & $-0.031$ & $-0.032$ & $-0.034$ &
& & & \\
\hline\hline
\end{tabular}%
}
\end{table*}

The on-site energies and hopping parameters are listed in Table~\ref{tab:hopping}. Under 5\% compressive isotropic strain, the dominant hopping parameters increase relative to ambient pressure. In particular, the nearest-neighbor in-plane hopping $|t_{[1,0]}^{x,I}|$ increases from 0.462 eV at ambient pressure to 0.548 eV at 5\% isotropic strain, while the vertical Inner--Inner $d_{z^2}$ hopping $|t_{[0,0]}^{z,II}|$ increases from 0.544 to 0.722 eV. These values are larger than those of bilayer La$_3$Ni$_2$O$_7$ \cite{Luo2023Bilayer}, indicating a relatively weaker correlation in the quadlayer. The simultaneous enhancement of intralayer and interlayer hoppings is also consistent with the DFT-calculated broader Ni-$e_g$ bandwidth. In contrast, under 4\% uniaxial strain, the intralayer hoppings are essentially unchanged due to the fixed in-plane lattice constants, whereas the vertical $d_{z^2}$ interlayer hoppings increase selectively. Specifically, $|t_{[0,0]}^{z,II}|$ increases from 0.544 to 0.639 eV and $|t_{[0,0]}^{z,IO}|$ increases from 0.543 to 0.612 eV. This orbital-selective enhancement of $d_{z^2}$ interlayer coupling could enhance interlayer superexchange coupling, which is considered to be favorable for superconductivity in RP nickelates \cite{Luo2023Bilayer, Wu2024SCPMA, Chen2024Trilayer, Sakakibara2024PRB, Yang2024TrilayerModel, Yang2023InterlayerValenceBonds, Lu2024PRL_Interlayer, Shen2023CPL, Qu2024BilayerTJ, Luo2024BilayerTJ, Qiu2026FilmPairing, Zhong2026FilmChargeTransfer}.

\begin{figure}[htbp]
\centering
\includegraphics[width=\columnwidth]{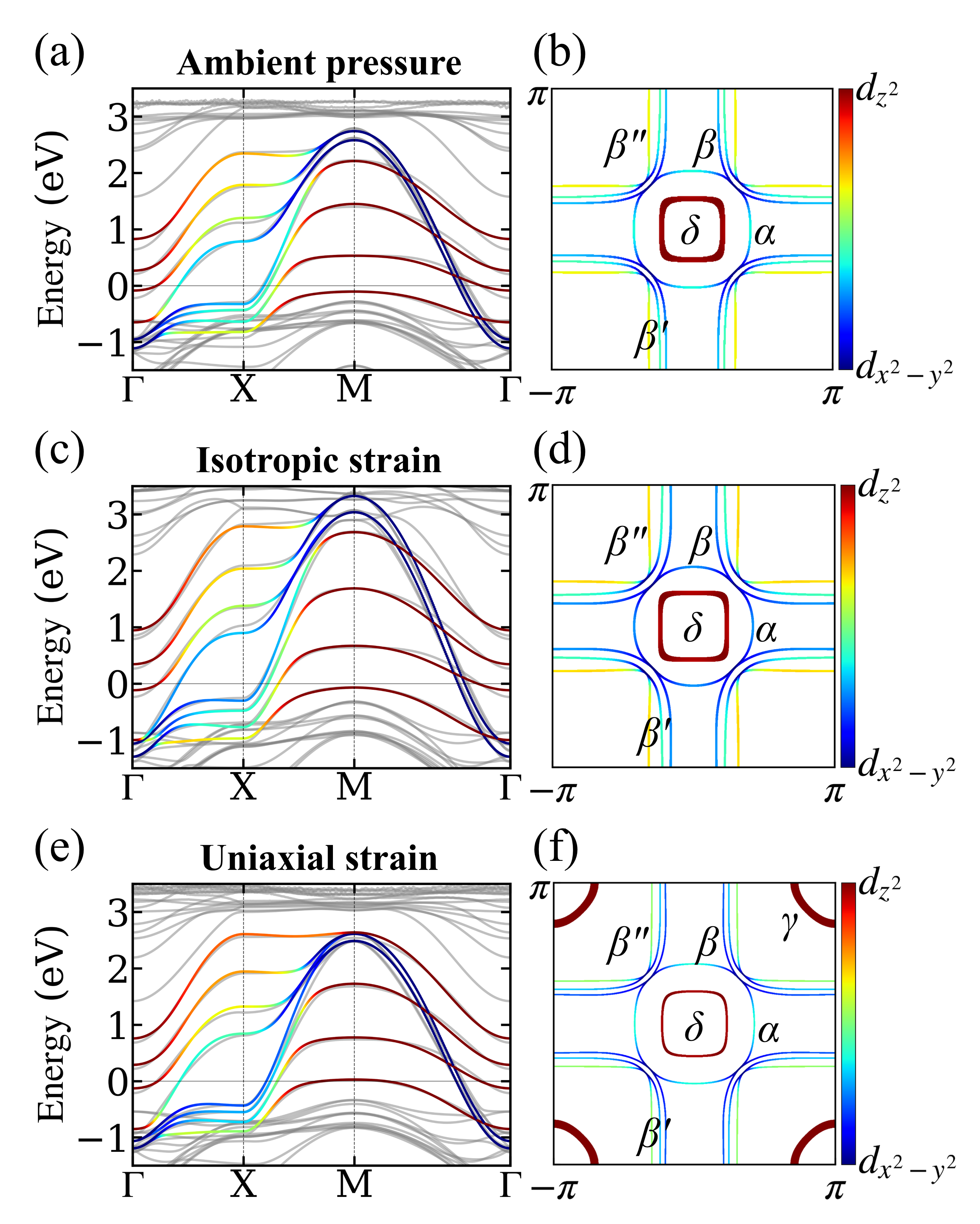}
\caption{\label{fig:tb_model} Tight-binding band structures compared with the DFT results (gray lines) and the corresponding Fermi surfaces at ambient pressure (a, b), $5\%$ isotropic strain (c, d), and $4\%$ $c$-axis uniaxial strain (e, f). The color scale indicates the orbital weight.}
\end{figure}

The resulting tight-binding band structures and Fermi surfaces are displayed in Fig.~\ref{fig:tb_model}. The low-energy Ni-$e_g$ dispersions obtained from DFT are well reproduced. At ambient pressure, the Fermi surface consists of five sheets ($\alpha$, $\beta$, $\beta'$, $\beta''$, and $\delta$), with the $\alpha$ pocket predominantly of $d_{x^2-y^2}$ character, and the $\Gamma$-centered $\delta$ pocket derived mainly from the $d_{z^2}$ orbital. Under 5\% isotropic strain, the Fermi surface remains similar, except for a slight expansion of the $\delta$ pocket. Notably, the $d_{z^2}$-derived $\gamma$ hole pocket around the $M$ point, which is present in the high-pressure bilayer $\mathrm{La}_3\mathrm{Ni}_2\mathrm{O}_7$ \cite{Luo2023Bilayer} and trilayer $\mathrm{La}_4\mathrm{Ni}_3\mathrm{O}_{10}$ \cite{Chen2024Trilayer}, is absent under both ambient pressure and $5\%$ isotropic strain, consistent with previous calculations \cite{Zhang2025Trends}. This indicates that purely hydrostatic compression is insufficient to induce the bonding1 $d_{z^2}$ band to cross the Fermi level in the quadlayer La$_5$Ni$_4$O$_{13}$.

However, 4\% uniaxial compressive strain brings a qualitative difference. The $d_{z^2}$-dominated band near the M point shifts upward and crosses the Fermi level, giving rise to an additional $\gamma$ hole pocket that is absent in the other two cases. The appearance of the $\gamma$ pocket is a direct consequence of uniaxial strain, which compresses the $\mathrm{NiO}_6$ octahedron along the $c$-axis and raises the energy level of $d_{z^2}$ orbital. This strain-induced Lifshitz transition distinguishes the effect of uniaxial strain from both ambient pressure and isotropic strain, and identifies uniaxial compression as a more efficient route than hydrostatic pressure for generating the $\gamma$ pocket in La$_5$Ni$_4$O$_{13}$.

To analyze the mirror-parity structure of the quadlayer model, we perform a unitary transformation of the layer basis using the mirror symmetry of the lattice \cite{Jung2022HigherOrder}. For a given orbital $\gamma \in \{x,z\}$, the original layer basis
\[
\Psi_{\gamma}=(c_{\gamma1},c_{\gamma2},c_{\gamma3},c_{\gamma4})^{\mathrm{T}}
\]
is transformed into the parity-adapted basis \cite{Chen2024Trilayer}:
\begin{equation}
\Phi_{\gamma}=(O^+,I^+,O^-,I^-)^{\mathrm{T}},
\end{equation}
where the mirror-even ($+$) and mirror-odd ($-$) combinations of the Outer ($O$) and Inner ($I$) layers are defined as
\begin{equation}
O^\pm=\frac{1}{\sqrt{2}}(c_{\gamma1}\pm c_{\gamma4}),
\quad
I^\pm=\frac{1}{\sqrt{2}}(c_{\gamma2}\pm c_{\gamma3}).
\end{equation}

Along the nodal directions $|k_x|=|k_y|$, the $d_{x^2-y^2}$ and $d_{z^2}$ sectors are decoupled since the inter-orbital hybridization terms vanish ($H^{xz}(\mathbf{k})=0$). Using the symmetry relations
$H_{11}^\gamma=H_{44}^\gamma$,
$H_{22}^\gamma=H_{33}^\gamma$,
$H_{12}^\gamma=H_{34}^\gamma$, and
$H_{13}^\gamma=H_{24}^\gamma$,
the intra-orbital Hamiltonian in the parity basis becomes block diagonal:
\begin{equation}
H'^\gamma(\mathbf{k})=
\begin{bmatrix}
H_{11}^\gamma+H_{14}^\gamma & H_{12}^\gamma+H_{13}^\gamma & 0 & 0 \\
H_{12}^\gamma+H_{13}^\gamma & H_{22}^\gamma+H_{23}^\gamma & 0 & 0 \\
0 & 0 & H_{11}^\gamma-H_{14}^\gamma & H_{12}^\gamma-H_{13}^\gamma \\
0 & 0 & H_{12}^\gamma-H_{13}^\gamma & H_{22}^\gamma-H_{23}^\gamma
\end{bmatrix}.
\end{equation}
The upper $2\times2$ block describes the sector spanned by $(O^+,I^+)$, whereas the lower $2\times2$ block describes the sector spanned by $(O^-,I^-)$. Each sector contains two eigenstates arising from the mixing between the corresponding Outer- and Inner-layer combinations. Within the $d_{z^2}$ sector, the bonding1 and antibonding1 states belong to the mirror-even sector and are formed by different linear combinations of $O^+$ and $I^+$, whereas the bonding2 and antibonding2 states belong to the mirror-odd sector and are formed from $O^-$ and $I^-$.


\subsection{Spin susceptibility}

To investigate the strain dependence of spin fluctuations in the quadlayer nickelate, we calculate the static RPA spin susceptibility $\hat{\chi}_{S}^{\mathrm{RPA}}(\mathbf{q},0)$ for the three two-orbital models using $U=1$~eV and Hund's coupling $J_H=U/6$. The same interaction parameters below the  respective RPA critical value $U_c$ are used for all three conditions, following the standard multiorbital RPA framework \cite{Kuroki2008, Graser2009} and related calculations for layered nickelates \cite{Luo2023Bilayer, Lechermann2023Correlations, Chen2024Trilayer, Zhang2025Trends, Hu2025FilmModel, Zhang2026HybridRP, Chen2026_1313}. At each momentum, the largest eigenvalue $\Lambda_{\max}(\mathbf{q})$ of the susceptibility matrix $\hat{\chi}_{S}^{\mathrm{RPA}}(pq,st)$ gives the spin susceptibility in the strongest channel \cite{Zhang2024PressurizedTrilayer}.

\begin{figure*}[!t]
\centering
\includegraphics[width=\textwidth]{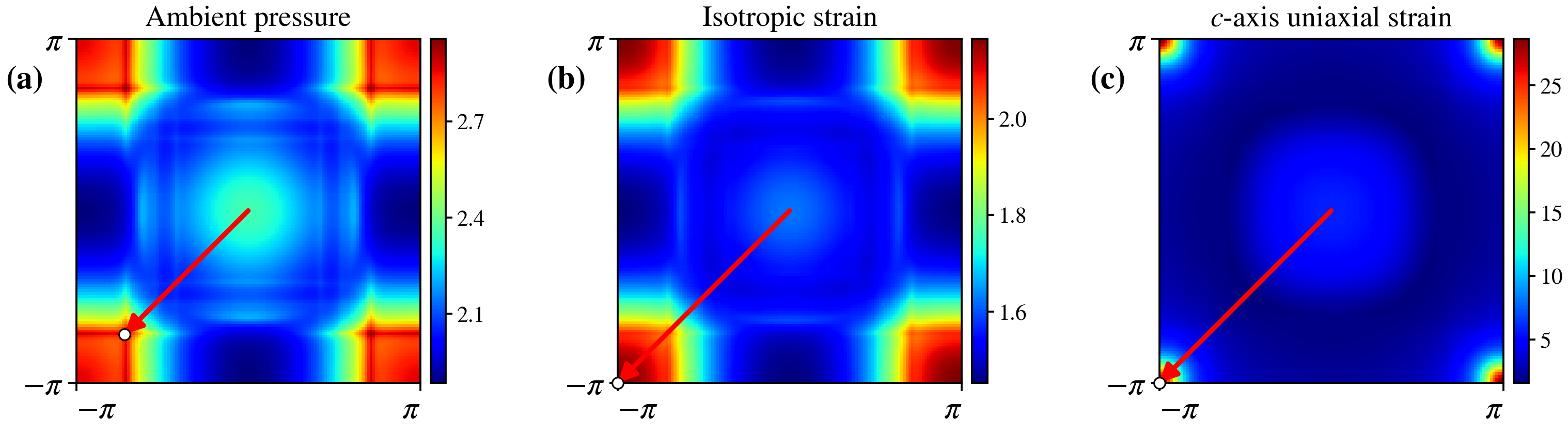}
\caption{\label{fig:rpa_eigen}
Momentum dependence of the largest eigenvalue $\Lambda_{\max}(\mathbf{q})$ of the static RPA spin susceptibility for (a) ambient pressure, (b) $5\%$ isotropic strain, and (c) $4\%$ $c$-axis uniaxial strain, calculated with $U=1$~eV and $J_H=U/6$. White circles mark representative maxima, and red arrows indicate the corresponding nesting vectors.}
\end{figure*}
Figures~\ref{fig:rpa_eigen}(a)--\ref{fig:rpa_eigen}(c) show the momentum dependence of $\Lambda_{\max}(\mathbf{q})$ at  $U=1$~eV. At ambient pressure, the susceptibility peaks near $\mathbf{q}=(2\pi/3,2\pi/3)$. Under both isotropic and $c$-axis uniaxial strain, the peak is located at $M$, $\mathbf{q}=(\pi,\pi)$. The peak value decreases under isotropic strain and increases substantially under $c$-axis compression. Part of the enhancement under uniaxial strain arises because this model is closer to its RPA magnetic instability at the chosen interaction strength.


Under isotropic strain, the Fermi surface remains similar to that at ambient pressure [Figs.~\ref{fig:tb_model}(b) and \ref{fig:tb_model}(d)], while the $d_{z^2}$-derived band top at $M$ lies just below the Fermi level [Fig.~\ref{fig:tb_model}(c)]. The susceptibility peak at $M$ suggests a possible contribution from ``virtual'' scattering between the incipient $\gamma$ band and the $\Gamma$-centered $\delta$ pocket, even though the $\gamma$ pocket has not formed yet. A similar mechanism has been discussed for the trilayer RP nickelate $\mathrm{La}_4\mathrm{Ni}_3\mathrm{O}_{10}$ \cite{Zhang2024PressurizedTrilayer}.

Under $c$-axis uniaxial strain, the upward shift of the $d_{z^2}$-derived band produces the $\gamma$ pocket at $M$ [Figs.~\ref{fig:tb_model}(e) and \ref{fig:tb_model}(f)]. The \(M\)-point susceptibility peak is consistent with scattering between the $\Gamma$-centered $\delta$ pocket and the $M$-centered $\gamma$ pocket. The appearance of the $\gamma$ pocket is accompanied by the enhanced susceptibility at $M$ in Fig.~\ref{fig:rpa_eigen}(c). The formation of a $d_{z^2}$-derived pocket and enhanced spin fluctuations have also been discussed in theories of superconductivity in bilayer and trilayer nickelates \cite{Luo2023Bilayer, Yang2023PRB, Liu2023PRL, Zhang2024BilayerRPA, Chen2024Trilayer, Yang2024TrilayerModel, Zhang2024TrilayerRPA, Zhang2024PressurizedTrilayer}. These similarities suggest that $c$-axis compression may promote a $\delta$--$\gamma$ scattering channel analogous to that favoring sign-changing $s^\pm$ pairing in bilayer and trilayer nickelates \cite{Zhang2024TrilayerRPA,Zhang2024PressurizedTrilayer}, providing a possible route to superconductivity in the quadlayer system.



\section{Conclusion}

In summary, we performed comprehensive DFT calculations on the quadlayer RP nickelate La$_5$Ni$_4$O$_{13}$ under ambient pressure, 5\% compressive isotropic strain, and $4\%$ $c$-axis uniaxial strain. DFT results show that isotropic strain broadens the Ni-$e_g$ bands and induces charge transfer from O-$p$ to Ni-$d$ orbitals. In contrast, $c$-axis uniaxial strain acts selectively on the $e_g$ orbitals, which leaves the Ni-$d_{x^2-y^2}$ dispersion nearly unchanged but shifts the bonding1 band upwards, accompanied by an increased Ni-$d_{x^2-y^2}$ occupation and a reduced $d_{z^2}$ occupation.

From Wannier downfolding, we constructed a quadlayer two-orbital model for the Ni-$e_g$ states. The two-orbital model reveals that isotropic strain enhances both in-plane and interlayer hoppings, while uniaxial strain selectively enhances the vertical $d_{z^2}$ interlayer couplings. Consequently, the $\gamma$ pocket remains absent under ambient pressure and isotropic strain, but emerges around the M point under 4\% uniaxial strain, signaling a strain-induced Lifshitz transition. This indicates that uniaxial compression is a more efficient route than hydrostatic pressure for generating the $d_{z^2}$-derived $\gamma$ pocket.

Based on the constructed models, we calculated the static RPA spin susceptibility at $U=1$ eV for all three structures. At the chosen interaction strength, the largest eigenvalue peaks near $\mathbf{q}=(2\pi/3,2\pi/3)$ at ambient pressure and at $M$ under both strain conditions. Under isotropic strain, the $M$-point peak arises from low-energy interband scattering involving the incipient $\gamma$ band. 
Under $c$-axis compression, the formation of the $\gamma$ pocket is accompanied by an enhanced spin susceptibility at $M$, consistent with a contribution from $\delta$--$\gamma$ inter-pocket scattering. These electronic and magnetic features resemble those discussed in theories of bilayer and trilayer nickelates \cite{Luo2023Bilayer, Chen2024Trilayer, Yang2023PRB, Liu2023PRL, Zhang2024BilayerRPA, Yang2024TrilayerModel, Zhang2024TrilayerRPA, Zhang2024PressurizedTrilayer}, suggesting that $c$-axis compression may provide favorable conditions for superconductivity in the quadlayer nickelate.

\begin{acknowledgments}
This work was supported by the National Natural Science Foundation of China (Grant Nos. 12494591, and 92565303), the National Key Research and Development Program of China (Grant No. 2022YFA1402802), the Guangdong Provincial Quantum Science Strategic Initiative (Grant No. GDZX2401010, GDZX2601002), the Guangdong Provincial Key Laboratory of Magnetoelectric Physics and Devices (Grant No. 2022B1212010008), and the Research Center for Magnetoelectric Physics of Guangdong Province (Grant No. 2024B0303390001). We also acknowledge the Dali Nickelate Forum.

\end{acknowledgments}
\bibliography{apssamp}

\end{document}